\documentclass{article}

\usepackage{arxiv}

\usepackage[utf8]{inputenc} % allow utf-8 input
\usepackage[T1]{fontenc}    % use 8-bit T1 fonts
\usepackage{hyperref}       % hyperlinks
\usepackage{url}            % simple URL typesetting
\usepackage{booktabs}       % professional-quality tables
\usepackage{amsfonts}       % blackboard math symbols
\usepackage{nicefrac}       % compact symbols for 1/2, etc.
\usepackage{microtype}      % microtypography
\usepackage{lipsum}
\usepackage{multirow}
\usepackage{graphicx}
\usepackage{float}
\usepackage{adjustbox}
\usepackage{array}
\newcolumntype{x}[1]{>{\centering\arraybackslash\hspace{0pt}}p{#1}}
\newcolumntype{L}[1]{>{\raggedright\let\newline\\\arraybackslash\hspace{0pt}}p{#1}}
\newcolumntype{C}[1]{>{\centering\let\newline\\\arraybackslash\hspace{0pt}}m{#1}}
\newcolumntype{R}[1]{>{\raggedleft\let\newline\\\arraybackslash\hspace{0pt}}p{#1}}

\title{Balancing Innovation and Integrity: AI Integration in Liberal Arts College Administration}

\author{
 Ian Read \\
  Soka University of America\\
  1 University Drive\\
  Aliso Viejo, CA 92656\\
  \texttt{iread@soka.edu} \\
}

\begin{document}
\maketitle
\begin{abstract}
This paper explores the intersection of artificial intelligence and higher education administration, focusing on liberal arts colleges (LACs). It examines AI's opportunities and challenges in academic and student affairs, legal compliance, and accreditation processes while also addressing the ethical considerations of AI deployment in mission-driven institutions. Considering AI’s value pluralism and potential allocative or representational harms caused by algorithmic bias, LACs must ensure AI aligns with its mission and principles. The study highlights other strategies for responsible AI integration, balancing innovation with institutional values.
\end{abstract}

% keywords can be removed

\section{Introduction}

Integrating artificial intelligence into higher education presents a vivid intersection of rapidly changing technology within society, eliciting predictions of terrifying ruin or breathtaking advancement. Liberal arts colleges (LACs), with their distinct missions and close-knit communities, stand at the crossroads of embracing or rejecting AI's transformative potential while safeguarding their deeply human-centered values. These universities can serve as a natural laboratory for AI integration, where emerging technologies can be carefully piloted, ethically evaluated, and continuously refined before broader adoption. Their interdisciplinary approach, faculty-led governance, and commitment to personalized learning make them ideal environments for testing and refining AI applications that prioritize human-centered values \cite{lang1999a}. Unlike large research universities that may implement AI at scale with limited oversight, LACs can experiment in controlled settings, ensuring ethical and educational priorities remain central.

Liberal arts colleges (LACs) balance intellectual engagement with social responsibility. Here also lies the crux: with fewer resources than large research universities, they must navigate the ethical and operational risks of AI—systems that are far from neutral as these institutions grapple with relatively fewer resources than large research institutions, they must carefully navigate the ethical and operational risks of AI systems, which are anything but neutral \cite{friedler2016a}. The concepts of fairness and harm, central to this navigation, reveal themselves to be fluid, contested, and deeply tied to the historical, cultural, and local contexts in which they are debated \cite{dewey1929a}.

Consider one important task laden with competing ideas of fairness: budget allocation. Universities face the delicate task of distributing resources in ways that reflect both their intrinsic mission and external pressures, such as market demand (e.g., student enrollment and grant funding potential). This process often involves navigating entrenched interests and institutional inertia, revealing the complexities of defining and achieving fairness \cite{massy1996a}. When one popular AI platform was tasked with distributing \$100 million separately to three hypothetical liberal arts colleges, it laid bare its implicit assumptions. For a generic liberal arts college, it ostensibly “balanced” equity and academic excellence. Yet, given the distinct contexts of a progressive, globally oriented West Coast college and a conservative, Christian-driven institution in the South, the AI’s allocations shifted dramatically. At the West Coast institution, cosmopolitanism and sustainability recategorized hundreds of thousands of dollars, while at the Southeastern college, Christian faith-based instruction and traditional community values rose to the fore \footnote{ChatGPT, response to “Create a budget that allocates \$100 million dollars across the various departments of a typical liberal arts college. Explain each allocation.” OpenAI, December 17, 2025; ChatGPT, response to “Create a budget that allocates \$100 million dollars across the various departments of Conservative, Christian-driven LAC in the South. Explain each allocation.” OpenAI, December 17, 2025; ChatGPT, response to “Create a budget that allocates \$100 million dollars across the various departments of a progressive, globally oriented LAC on the West Coast. Explain each allocation.” OpenAI, December 17, 2025.}. These variations illustrate that AI can adapt to values when provided with context \cite{zhou2023a}. Absent such guidance, it defaults to varying median values that depend on its training data and must overlook the unique character of any given institution or community.  It is no surprise, then, that different AI tools produce hugely varying outcomes due to distinct training data, languages, and algorithms \cite{uelgen2025a}. 

The budget exercise reveals what experts in computer science widely call the “alignment problem,” or the challenge of aligning AI with human values, preferences, or needs \footnote{“How to ensure that these [AI] models capture our norms and values, understand what we mean or intend, and, above all, do what we want — has emerged as one of the most central and most urgent scientific questions in the field of computer science. It has a name: the alignment problem” Brian Christian, The Alignment Problem: Machine Learning and Human Values (New York: W.W. Norton \& Company, 2020), 12.}. Despite the many attempts, “alignment” is a Sisyphean effort, for it disregards value pluralism for a monistic ideal \cite{rudschies2020a,mishra2023alignment,elmahjub2023ai}. As Shannon Vallor writes, “AI isn't developing in harmful ways today because it's misaligned with our current values. It's already expressing those values all too well”\cite{vallor2024a}. Vallor’s point is that humans themselves never align or share values, at least in large groups, and a millennium-old history of grappling with fairness and justice has not solved this “problem,” nor would we want the world’s value diversity flattened by anything capable of complex and universal decision-making. Fortunately, there is much wisdom in the many philosophical approaches to value conflicts, from deontological principles, which emphasize universal duties, to relational approaches like care ethics, which focus on interpersonal context, to pragmatism, which emphasizes practical consequences and adaptability in ethical decision-making \cite{sandel2009a}. As value alignment has shown itself impossible beyond large groups, it’s unsurprising that efforts have increasingly turned toward mitigating harm through diverse and often competing regulations and oversight, leading to a rapidly growing patchwork of government and corporate AI guidelines and regulations \cite{gabriel2020a,white-a}.

As the focus shifts from adhering to universal principles to mitigating harm with a patchwork of national, regional, and community-based rules, ensuring the least harmful integration of AI systems into academic and administrative practices becomes essential \cite{association2024a,dotan2024a}. Mitigation in this context can be analyzed through two primary lenses: allocative harm and representational harm \cite{barocas2019a}. Allocative harms occur when opportunities or resources are withheld from certain groups, such as when algorithms determine admission or job offers. These harms can be easier to measure, but their recognition as problematic often depends on ethical approaches to fairness, although some are clearly defined in law. Representational harm can also be subtle or disputed and occurs when systems stigmatize or stereotype groups, as seen in language models that encode and perpetuate stereotypes \cite{peters2023a,chien2024a}. For example, an AI model used in university admissions might avoid representational harm by removing demographic indicators like race or gender from its training data \footnote{Even in this case, other variables like ZIP code or school district might become proxies for race, thereby still encoding demographic patterns, leading to representational harm.}. However, this could lead to allocative harm if it overlooks systemic inequalities, such as disparities in access to advanced coursework or extracurricular opportunities, effectively disadvantaging underrepresented groups.

These challenges mirror those faced by human decision-makers, raising the question of whether AI might ultimately perform better in specific contexts. Humans often encounter similar difficulties, as personal biases, lack of calibration, and opaque reasoning can lead to comparable harms \cite{kahneman1979a,martino2006a}. Machines may have certain advantages, such as ensuring calibration and consistency in decisions, although there is a long history of misplacing this hope in machines \cite{bates2024a}. Setting aside that algorithms might be consistently unfair, both humans and AI systems can perpetuate a third important type of harm—procedural—such as a lack of transparency in how decisions are made or the inability of affected individuals to challenge or appeal outcomes \cite{hagendorff2023a,decker2024a}. In human contexts, procedural harm might arise from hasty, informal, or opaque decisions, while in AI systems, it often stems from complex algorithms that use the “wrong” or conflicting definitions of fairness or are challenging to interpret \cite{kwoka2022a}.

Liberal arts colleges should critically examine these trade-offs, evaluating AI systems for fairness and comparing them to human abilities and limitations while always keeping humans in the loop. Addressing the three dimensions of harm—allocative, representational, and procedural—requires moving beyond vague accusations of bias or lofty claims of solving the “alignment problem,” such as through ever more finely tuned algorithms.	  Since human and AI decision-making effectiveness depends on how well these processes are designed and deployed, efforts should ensure robust oversight and make both human and AI systems transparent and open to scrutiny, particularly as they evolve into community or domain-specific systems. 

When we set aside vague “bias” as the problem and monistic alignment as a goal, we can work with rather than attempt to eliminate AI’s many inherent and ultimately unresolvable value contradictions \cite{samuel2022a}. One prominent example is the dominance of WEIRD (Western, Educated, Industrialized, Rich, and Democratic) perspectives in AI training data, which skews outputs to reflect only a subset of global human experiences \cite{johnson2022a}.  Most widely used AI models, particularly those developed in the U.S. and Europe, are trained on datasets predominantly sourced from English-speaking Western countries, inherently privileging perspectives from a fraction of the world’s nearly 8 billion people. This lack of global representation leads to recommendations imbued with embedded value systems—intentionally programmed or implicitly inherited—that may fail to account for diverse cultural norms, communal values, or non-Western priorities. Efforts to mitigate this problem, such as developing localized AI systems tailored to regional contexts, are rising. For instance, China has initiated projects to create models aligned with collectivist cultural norms, while other nations emphasize multilingual datasets to capture a broader range of perspectives \cite{lucas2023china,clark2023a,india2022bhasha}. However, addressing WEIRD values through inclusivity introduces its own challenges, such as reconciling conflicting value systems—for example, balancing freedom of speech with religious traditions—within any single AI system. The growing trend toward fragmentation, or “personalized” GPTs, again underscores the impossibility of universal alignment while emphasizing the need for adaptable regulations to maximize benefits and mitigate harm from value-laden systems, including at the university community levels.

To develop strategies for the ethical and pragmatic implementation of AI systems, universities must also participate in their own critical examination of how principles like safety, fairness, privacy, and transparency are conceptualized and applied. Few organizations are better equipped to draw on and consider frameworks' philosophical, ethical, and practical application within purpose-driven communities than liberal arts colleges. Achieving this will require significant and sustained effort, such as risk assessment, iterative testing, stakeholder engagement, and balancing trade-offs, a tall order for many resource-constrained colleges.

Because of value heterogeneity, we already see a proliferation of AI systems tailored to distinct contexts and individual or organizational priorities \cite{marquis2024a,elmahjub2023ai}. Companies and universities, at least those with the resources, are creating their own GPTs or internal AI systems, “fine-tuned” with proprietary data and prompts.  Among the thousands of universities without their own systems, faculty, staff, and students are turning to an increasing array of AI tools, with hundreds of products on the market \cite{mckinsey2023a,research2024a}. Recent data indicates a substantial rise in AI utilization among higher education professionals, with 84\% reporting usage in their professional or personal lives—a 32 percent increase over the past year \cite{ellucian-a}. Additionally, 74\% of presidents and chancellors polled by The Higher Learning Commission, a university accreditation body, report implementing generative AI technologies within their institutions.\cite{commission-b}  Considering that AI is also integrated into internet searches on browsers like Chrome or Edge, with AI results often presented first, we might accurately say that nearly everyone now uses AI. ChatGPT remains the most popular at the time of writing, but several nearly equally performing competitors have joined it \cite{capitalist-a,analysis-a}.

This essay has mostly discussed generative AI (GenAI). In fact, GenAI is only one of several types of AI, and we should distinguish it from predictive AI (PAI) and “narrow AI.” GenAI creates new content or data that resembles patterns in its training data, using underlying architectures such as transformer-based models \cite{vaswani2017a}. For instance, a GenAI system might simulate budgetary recommendations or craft hypothetical scenarios tailored to specific institutional contexts by drawing on large datasets of textual (e.g., large language model or LLM) or numerical information \cite{yenduri2023a}. In contrast, predictive AI focuses on analyzing historical data to forecast future outcomes or trends. PAI typically employs statistical models or machine learning techniques, such as regression analysis or decision trees, to provide actionable insights \cite{i2023a}. While GenAI synthesizes new possibilities based on learned patterns, PAI identifies relationships within structured data to project-specific probabilities or outcomes, such as predicting student retention rates or enrollment trends. Finally, narrow AI refers to tools designed for specific, restricted applications, such as grammar checks or automated scheduling, whose functionality is tightly constrained to a particular task.  Despite their differences, all types of AI share the use of algorithms to generate decisions or outputs.  Additionally, all AIs carry unique strengths and limitations, and none escape humanity’s extraordinarily varied and contradictory value systems. Their differences underscore the importance of selecting the right tool \cite{harrington2024a}.

The rapid proliferation of AI tools risks amplifying institutional disparities, complicating interoperability, and scattering governance, legality, and accountability across uncoordinated systems \cite{selwyn2019a}. While wealthier institutions may navigate these challenges more effectively, economically struggling liberal arts colleges (LACs) face heightened vulnerability, compounded by broader economic and demographic shifts threatening their financial viability. Rising tuition costs and mounting student debt—reaching \$1.75 trillion in 2024—have fueled growing skepticism about the value of a college degree, while a 1.4\% decline in the U.S. college-age population between 2010 and 2020 has contributed to enrollment drops \cite{lendingtree-a,bureau2021a}. Nearly 300 colleges and universities offering an associate degree or higher closed between 2008 and 2023, with over 60\% of these being for-profit institutions \cite{report-a}. LACs, reliant on tuition as their primary funding source and burdened by the high per-student costs of small class sizes and personalized education, are particularly at risk of closure.\cite{capstone2025liberal} To survive, many will turn toward artificial intelligence to streamline operations, reduce inefficiencies, and enhance their mission of fostering ethically grounded, adaptable graduates. Falling costs and the increasing availability of efficient open-source AI systems, such as LLaMA or DeepSeek-V3, present an opportunity for LACs to adopt tailored AI solutions, as these may lower barriers to implementing proprietary or open-source models designed to meet their specific needs.

In just a few years, most colleges will likely use their own “fine-tuned” large language models (LLMs) tailored to their specific needs and values. This shift will reflect a sharp fall in the barriers to AI adoption and bring new challenges, including ensuring these systems align with institutional missions, mitigate harm, and comply with legal and ethical standards. For liberal arts colleges (LACs), this presents a transformative opportunity to enhance student support, enrich educational experiences, and demonstrate leadership in ethical AI integration. This essay argues that LACs, with their interdisciplinary focus and smaller scale, are uniquely positioned to set an example for higher education and other industries by balancing technological innovation with their mission to foster holistic education and community values. To support this argument, this essay examines the role of AI in academic and student affairs, explores the legal and ethical risks associated with AI tools, and highlights the role of accreditation agencies and faculty and staff training in tailoring these systems to reflect institutional values, cause no harm, and uphold the law. By addressing these challenges with intentionality and foresight, LACs can position themselves as adopters of AI and leaders in shaping its responsible use across higher education.

\section{AI in Academic Affairs}

In the best liberal arts colleges, the air hums with intellectual curiosity, ethical reflection, and a kind of cognitive dexterity—the ability to weave threads from disparate disciplines into something rich and meaningful. These institutions nurture critical thinking, not as a skill to be ticked off a checklist, but as a way of being: questioning assumptions, embracing diverse perspectives, and tackling problems with evidence and rigor. Contrast this with the passivity that can settle in when information is absorbed without question and authority accepted without scrutiny. The best liberal arts colleges stand apart by creating spaces where small class sizes foster real conversation, mentorship, and moments of personal discovery—habits of mind that form the bedrock of transformative learning \cite{kuh2008a}.

Enter artificial intelligence, a technology that promises to reshape the scaffolding of these institutions. In academic affairs, AI offers efficiency—streamlining course schedules, automating curriculum management, and liberating faculty to focus more on engaging with students. But there’s a catch. When AI is adopted piecemeal, fragmented systems can undermine the integrated, interdisciplinary ethos that defines the liberal arts. To preserve their mission, these colleges need not just technology but a strategy, one that thoughtfully aligns AI tools with institutional values.

GenAI, for example, can simulate debates, present complex scenarios, and offer challenges tailored to push students beyond their comfort zones \cite{education2023a}. AI tools can encourage students to grapple with conflicting evidence by generating arguments from multiple perspectives, nudging them toward deeper understanding \cite{lodzikowski2024generative}. Adaptive systems can cater to individual learning needs, and prompting interdisciplinary AI frameworks might help students see connections between fields they might otherwise overlook. But here, too, the risks loom large. Relying too much on AI can turn inquiry into rote acceptance, allowing the tools meant to foster curiosity to erode it instead. And let’s not forget the value pluralism and contradictions baked into algorithms or the blind spots where human diversity is flattened into something sterile and generic.

At its core, the debate over GenAI in universities is about more than gadgets and code. It’s about whether technology can genuinely enhance the personalized, creative, and critical experiences that define education—or whether it will serve as a Trojan horse for diminished rigor and deepened inequities. Yes, AI can personalize learning, automate the repetitive, and spark creativity. But it also risks plagiarism, unequal access, and, perhaps most alarmingly, a loss of the human connection that makes education transformational \cite{steponenaite2023plagiarism}. The task ahead is daunting but straightforward: to embrace innovation without losing sight of the practices that make education a profoundly human endeavor \cite{education2023a,memarian2023a,kasneci2023a,samala2024a}. In this balancing act lies the future of higher education.

This essay does not aim to resolve the many debates on GenAI’s utility and risk in classrooms. Instead, it shifts focus to a topic receiving far less attention despite its importance: how AI tools are beginning to improve efficiencies in administrative tasks within the university’s largest and most important areas. Regardless of whether one believes AI will help or harm classroom teaching and student learning, most agree that professors teach and mentor better when they have more time for students. Especially in liberal arts colleges, where close faculty-student relationships are central to the mission, freeing up faculty time can significantly enhance teaching and learning. However, as research highlights, while AI might streamline tasks often disproportionately undertaken by women and tenured faculty, such as service work, the risk remains that institutional expectations will rise alongside efficiency gains, potentially undermining the equity these tools aim to foster \cite{o2007a}. Institutions must, therefore, use AI thoughtfully to reduce inequities without increasing burdens.

AI is already transforming academic affairs, and the challenge will be maintaining the personalized, human-centered ethos of liberal arts education while pioneering advancements in efficiency and innovation \cite{delbanco2012a}. As shown in Table \ref{tab:1}, Academic affairs departments handle numerous repetitive processes, including course scheduling, student record audits, accreditation reporting, and curriculum planning. AI offers significant potential to automate these tasks, improving efficiency and enabling faculty to dedicate more time to their core responsibilities of teaching, mentoring, research, and fostering intellectual growth \cite{kuh2008a}. Predictive AI (PAI) can streamline scheduling by balancing faculty availability with student demand, while GenAI can automate drafting compliance reports and trend analyses. AI-powered tools also facilitate system integration, linking course data, learning management software (LMS) platforms, and advising records to reduce redundancies and enhance coherence. However, thoughtful implementation is critical to avoid pitfalls such as algorithmic biases and data privacy concerns, which could jeopardize the mission of liberal arts colleges \cite{mehrabi2021a}. 

\begin{table}[t]
\centering
\caption{AI tools within Academic Affairs}
\label{tab:1}
\begin{adjustbox}{max width=\linewidth}
\begin{tabular}{L{3cm}L{3cm}L{3cm}L{3cm}L{3cm}L{3cm}L{3cm}}
\toprule
\textbf{Task Group}                                                    & \textbf{Workflow Management}                                                            & \textbf{Scheduling and Resource   Allocation}                             & \textbf{Reporting and Dash-boards}                                                  & \textbf{Document Management}                                  & \textbf{Communications}                                         & \textbf{Data Integration}                                              \\
\midrule
Course scheduling,   catalog management, curriculum review             & Approving new courses,   tracking pre-requisites.                                        & Optimizing classroom   schedules, adjusting course times.        & Generating course demand   reports.                                       & Uploading updated syllabi   to central repositories. & Emailing schedule changes   to faculty.                 & Syncing course data with   ERP systems.                       \\
Accreditation   documentation, assessment cycles, compliance reporting & Routing accreditation updates,   scheduling assessments.                                 & Assigning staff to audit   cycles, syncing calendars.            & Generating annual   compliance reports.                                    & Archiving past compliance   reports.                  & Sending reminders for   accreditation deadlines.        & Integrating compliance   software with institutional records. \\
Faculty workload   balancing, evaluations, promotion and tenure        & Tracking evaluation   approvals, submitting promotion files.                             & Allocating committees for   evaluations, balancing workloads.   & Creating workload   comparison charts.                                     & Storing tenure review documentation.                  & Alerting faculty about   evaluation deadlines.          & Connecting evaluation   data with performance databases.      \\
Academic advising   sessions, degree audits, enrollment management     & Assigning students to   advisors, logging sessions.  Degree   requirement communications & Assigning advisors to new   students, redistributing caseloads. & Monitoring advisor   effectiveness reports. Academic “red-flag” reporting. & Archiving advising   records securely.                & Notifying students about   missing degree requirements. & Syncing advising   platforms with enrollment systems.         \\
Trend analysis, program   review, strategic planning                   & Routing program review   proposals.                                                      & Allocating reviewers for   institutional research proposals.     & Creating enrollment trend   charts.                                        & Archiving past strategic   plans.                     & Sending follow-ups about   data requests.               & Linking institutional   trends with external datasets.        \\
Data integration, LMS   analytics, academic technology support         & Updating LMS user permissions,   integrating software.                                  & Scheduling LMS system updates,   syncing external tools.         & Generating LMS usage   analytics.                                          & Storing LMS data backups.                             & Alerting users about LMS   outages.                     & Connecting LMS tools with   student information systems.      \\
\bottomrule
\end{tabular}
\end{adjustbox}
\end{table}

\subsection{Workflow Management}

Workflow management tasks, such as approving new courses, tracking prerequisites, and routing faculty or program reviews, are essential for maintaining the smooth operation of liberal arts colleges \cite{kuh2008a}. These processes often involve coordinating across multiple departments and heads of faculty and ensuring compliance with institutional policies. For example, routing accreditation updates requires managing deadlines and ensuring all necessary documentation is submitted. Tools like Trello and Asana can partially automate these workflows by providing task tracking and notifications but rely heavily on manual inputs \cite{techco-a}. PAI might enhance these processes by anticipating workflow bottlenecks or delays, while generative AI could use rubrics and standardized templates to streamline and improve course proposals or accreditation documents \cite{johnson2024regai}. This allows faculty and administrators to focus on more strategic aspects of academic planning rather than repetitive administrative tasks.

\subsection{Scheduling and Resource Allocation}

Scheduling and resource allocation tasks, such as optimizing classroom schedules, assigning staff to audit cycles, and redistributing advisor caseloads, are particularly challenging due to their complexity \cite{mohamed2016interactive}. Liberal arts colleges, with their small class sizes and personalized approaches, require tailored scheduling solutions that balance the needs of students, faculty, and facilities. For example, Coursedog offers integrated academic and event scheduling solutions, enabling course section planning, instructor assignments, and room bookings, with some bi-directional integrations for real-time updates with Student Information Systems (SIS) \cite{coursedog-a}.  Similarly, Accruent EMS Scheduling, widely used in higher education, supports academic and non-academic event scheduling with features like space optimization and conflict detection to centralize processes and reduce administrative burdens \cite{accruent-a}. While these systems improve operational efficiency, their ability to adapt dynamically to immediate changes remains limited, requiring careful configuration and ongoing manual oversight.  With the incorporation of AI, we can expect scheduling tools to dynamically adjust schedules based on real-time data, such as faculty availability or room usage trends, and provide predictive insights to optimize resource allocation. Additionally, these tools could generate customized scheduling scenarios or event recommendations, adapting to institutional needs with minimal manual intervention. By automating these intricate processes, colleges can free up resources for fostering more meaningful in-person interactions.

\subsection{Reporting and Dashboards}

Reporting and dashboard creation are key areas in which AI has significant potential to transform liberal arts colleges. Generating reports, such as annual compliance summaries or enrollment trend analyses, often involves labor-intensive extracting and manually compiling data from multiple sources. Tools like Power BI and Tableau already streamline this process by automating data visualization and configuration, but they increasingly integrate advanced AI features. For instance, Power BI incorporates AI Insights and automated machine learning to apply machine learning models for sentiment analysis, anomaly detection, and predictive analytics \cite{microsoft-a,microsoft-b}. Similarly, Tableau, with its integration of Salesforce’s Einstein AI, leverages GenAI to create visualizations, calculated fields, and tailored insights through conversational interfaces \cite{tableau-a,salesforce-a}. These capabilities can enhance trend identification, anomaly detection, and the generation of narrative summaries for accreditation or internal reviews. For example, AI could flag a decline in course demand within a discipline and suggest resource reallocation strategies based on historical patterns.  Predictive analytics systems, such as at Greenville University, may provide real-time academic risk assessments, allowing faculty to intervene earlier with students who show signs of disengagement or academic difficulty \cite{gregory2021a}. These tools empower administrators to prioritize decision-making and implement solutions rather than compile data by reducing manual input and focusing on actionable insights.

\subsection{Documentation and Record Management}

Documentation and record management tasks, such as storing tenure review files, archiving compliance reports, and maintaining advising records, are essential for operational continuity and compliance with state and federal regulations \cite{eaton2015a}. Tools like DocuWare and Laserfiche already help automate document storage and retrieval, offering features such as workflow automation and template-based tagging, though some manual setup is still required \cite{docuware2025education,laserfiche2025education,laserfiche2025summarization}. Integrating PAI into these systems could further enhance efficiency by detecting inconsistencies or identifying missing files, while GenAI could automate the creation of summaries or templates for frequently used documents \cite{chowdhury2024a}. For example, when preparing tenure documentation, AI could ensure all required materials are included and formatted correctly, saving time and improving compliance. It might be tailored to improve the accuracy of tenure reports, aligning the needs or priorities of review committees and administration with the faculty’s teaching, research, and service records. AI-driven solutions could also integrate with Student Information Systems (SIS) or Learning Management Systems (LMS), creating a unified data ecosystem to streamline administrative workflows further. 

\subsection{Communications and Notifications}

Effective communication is central to the mission of liberal arts colleges, and AI has the potential to make notifications and reminders more personalized and efficient. Sending reminders for accreditation deadlines or notifying students about missing degree requirements is essential but time-consuming. Tools like Mailchimp and Outlook Automations automate bulk communications but often lack the personalization needed for student-centered institutions. AI-powered features like Microsoft’s Co-Pilot in Outlook, Gemini in Gmail, and Intuit Assist in Mailchimp now offer generative AI capabilities to draft contextualized email responses or reminders, though these still require review and customization to ensure accuracy and alignment with institutional values \cite{microsoft2025copilot,google2025gemini,intuit2025mailchimp}. Additionally, Retrieval-Augmented Generation (RAG) could further enhance this by integrating institutional policies, communication guidelines, and prior examples into AI-generated drafts, ensuring consistency with the institution's values \cite{lewis2020a}. For instance, RAG could retrieve language emphasizing personalized education and inclusivity or flag potential privacy violations, such as improper sharing of student data, suggesting secure alternatives. However, users may feel spied on if the system over-monitors or uses overly intrusive techniques. To address this, transparency about how the AI generates its recommendations, and clear boundaries on data usage must be established, ensuring trust while aligning communications with the institution’s mission \cite{liu2021a}.

\subsection{System Integration and Data Syncing}

System integration and data syncing are critical for ensuring consistency and coherence across departments at liberal arts colleges. Tasks like syncing course data with Enterprise Resource Planning (ERP) systems like PeopleSoft or Workday, integrating compliance software with institutional records, and linking LMS tools with student information systems often require significant manual effort \cite{oracle-a,workday2025highereducation}. For example, updating a course catalog in the ERP system might involve separately entering the same data into the LMS and advising platform, increasing the risk of errors and inefficiencies. Tools like Zapier and MuleSoft provide robust automation options with advanced features, but their effectiveness in handling highly complex, real-time integrations may depend on extensive customization and proper configurations. Zapier connects different applications through prebuilt workflows called "Zaps," which trigger automated actions based on specific conditions. MuleSoft, on the other hand, operates as an integration platform that enables organizations to connect systems, applications, and data through APIs, facilitating more extensive and customizable integrations \cite{zapier-a,mulesoft2025integration}. 

While these tools may provide valuable automation options for integration, their reliance on predefined workflows and manual configuration highlights the need for more advanced solutions. Predictive and generative AI offers the potential to enhance these integrations by automating complex tasks, identifying inconsistencies, and providing actionable insights that existing tools may not fully address. Companies like IBM Watson Education and SnapLogic are already leveraging AI to streamline system integration, offering tools that automate documentation, generate connectors, and provide personalized support for universities \cite{ibm2025education,snaplogic-a}. Predictive AI could identify discrepancies across systems, while generative AI might suggest solutions or generate real-time data visualizations to improve decision-making. However, implementing AI-driven integrations involves significant costs, extensive staff training, and potential resistance to adoption—issues particularly pressing for smaller institutions with limited resources \cite{meeker2024a}. Moreover, the success of predictive and generative AI relies on access to high-quality, well-organized data, and these systems may struggle with nuanced or incomplete information. Despite these challenges, an AI system that seamlessly syncs changes in a course catalog with advising platforms and notifies advisors of relevant updates exemplifies how such integration can reduce redundancies and ensure faculty and administrators have accurate, up-to-date information for informed decision-making.

In the context of liberal arts colleges, integrating AI tools in academic affairs departments is not just an opportunity to enhance efficiency—it is a test of how well these technologies can align with the mission-driven values that define these institutions. While workflow optimization, scheduling, reporting, and system integration offer substantial potential to reduce administrative burdens, the fragmented adoption of AI tools risks creating silos and conflicting priorities. To truly serve the unique ethos of liberal arts colleges, AI implementation must be guided by a coordinated strategy that prioritizes interoperability and institutional coherence. This alignment ensures that operational improvements support, rather than detract from, the core educational goals of fostering critical thinking, collaboration, and personalized learning. By embedding their values into the deployment of AI systems, liberal arts colleges can lead the way in demonstrating how technology can complement, rather than compromise, the human-centered practices at the heart of education.

\section{AI in Student Affairs}

As AI weaves its way into university systems, student affairs departments find themselves at the intersection of promise and peril. These departments, the beating heart of campus life, oversee everything from mental health support to career counseling—domains rich with opportunities for innovation but fraught with potential risks. AI offers tantalizing prospects of automating scheduling, tracking data, and reducing administrative burdens, creating more space for meaningful human interaction. But alongside these efficiencies come formidable challenges: the risk of values contradictory to missions, compromising privacy, and diluting the personal connections that define the college experience. Liberal arts colleges, focusing on ethical reasoning and interdisciplinary collaboration, are uniquely positioned to navigate these tensions. Yet, the pressures they face—shrinking enrollments, economic uncertainties, and an ever-growing web of regulations—threaten to rush AI adoption in ways that could entrench disparities and weaken their human-centered missions.

AI offers significant opportunities to streamline repetitive administrative tasks such as scheduling, data tracking, and managing routine inquiries, freeing staff to focus on meaningful student interactions \cite{jacques2024a}. However, the benefits of these tools are often overstated, with companies downplaying risks such as algorithmic bias in housing assignments, inequities in job matching, and delays in mental health interventions due to misclassification \cite{eab2025ai}. Real-world examples demonstrate AI's potential and limitations: Penn State World Campus uses AI to streamline transfer credit evaluations, Maryville University automates transcript processing to reduce manual effort, and Kellogg Community College leverages an AI-powered CRM system to enhance communication efficiency \cite{brady2024a}. As Table \ref{tab:2} highlights, AI can assist with tasks like automating housing assignments, identifying at-risk students through predictive analytics, and managing event logistics, but these tools come with significant pitfalls, including privacy risks and system value conflicts. Balancing these efficiency gains with potential harms requires keeping human oversight at the center of AI implementation.

\begin{table}[t]
\centering
\caption{AI Tools and Risks in Student Affairs}
\label{tab:2}
\begin{adjustbox}{max width=\linewidth}
\begin{tabular}{L{2cm}L{4cm}L{4cm}L{4cm}L{4cm}}
\toprule
\textbf{Role}                      & \textbf{Tasks}                                                                                  & \textbf{Areas Where AI Might Best Serve}                            & \textbf{Areas with Greater Risk}                                                                                     & \textbf{Common Apps Used}                                      \\
\midrule
Director of Residence Life         & Managing housing assignments, Processing   maintenance requests, Resolving conflicts            & Automating housing assignments, Chatbots   for maintenance requests & Biased housing assignments (e.g., grouping   based on demographic data), Privacy risks from centralized housing data & StarRez, Roompact, AppFolio's Realm-X                          \\
Resident Hall Coordinator          & Overseeing residence halls, Tracking attendance at   programs, Supporting RAs                   & Attendance tracking, Automated reminders for events                 & Over-reliance on attendance data for engagement   metrics, Missing interpersonal nuances                             & Anthology (formerly Campus Labs) Engage, Eventbrite,   Scandit \\
Director of Student Activities     & Planning events, Coordinating budgets,   Supporting student organizations                       & Event logistics automation, Budget tracking   and approval systems  & Inequitable allocation of funds or event   access, Bias in engagement metrics                                        & Presence, Engage/Campus Labs                                   \\
Career Counselor/Coach             & Reviewing resumes, Matching students with job postings,   Hosting workshops                     & AI resume review, Job-matching algorithms                           & Bias in resume parsing or job recommendations (e.g.,   privileging traditional paths)                                & Handshake, Symplicity                                          \\
Internship Coordinator             & Finding internship opportunities, Managing   applications, Following up on placements           & AI for internship matching, Automated   follow-ups                  & Biased internship matching favoring   well-connected students                                                        & Symplicity, Handshake                                          \\
Director of Counseling Services    & Managing counseling appointments, Triage for student   mental health, Running wellness programs & AI scheduling assistants, Initial mental health triage   tools      & Misclassification of urgency in mental health needs,   Privacy risks from sensitive data                             & Titanium, Ivy.ai, Ocelot                                       \\
Mental Health Counselor            & Providing therapy, Crisis intervention,   Educating students on wellness                        & Chatbots for non-urgent FAQs, Follow-up   surveys                   & Failure to detect complex emotional needs   or nuances                                                               & Ivy.ai, Ocelot                                                 \\
Retention Coordinator              & Identifying at-risk students, Monitoring retention   data, Designing interventions              & AI analysis of retention trends, Early-warning systems              & False positives/negatives in identifying at-risk   students, Bias in predicting student success                      & Starfish, EAB Navigate                                         \\
Accessibility Services Coordinator & Managing accommodations, Processing   documentation, Educating staff on accessibility           & Workflow automation for accommodation   requests, Reminder systems  & Misinterpreting or deprioritizing nuanced   accessibility needs                                                      & Accommodate, Clockwork                                         \\
Director of Community Service      & Planning service-learning projects, Tracking volunteer   hours, Building community partnerships & Automating volunteer tracking, Event reminders                      & Over-reliance on metrics, Undervaluing informal service   contributions                                              & Presence                                                       \\
Director of Campus Recreation      & Organizing intramural sports, Managing   fitness facilities, Tracking participation             & AI scheduling for leagues/events,   Participation tracking          & Excluding students without digital access,   Bias in participation incentives                                        & IMLeagues, Fusion                                              \\
\bottomrule
\end{tabular}
\end{adjustbox}
\end{table}

As the table demonstrates, housing and residence life staff often rely on platforms like Roompact and StarRez to manage housing assignments and communication. StarRez integrates AI features, such as its "AI Email Assistant" for personalized communication, while Roompact has expressed skepticism about overreliance on AI through client interviews \cite{starrez-a,roompact2024ai}. Meanwhile, AppFolio Realm-X claims “revolutionary” AI functionality, allowing users to “ask general product questions, retrieve data from a database, streamline multi-step tasks, and automate repetitive workflows in natural language without an instruction manual.”\cite{appfolio2023a} Fyma leverages AI-powered computer vision through existing CCTV systems to analyze space utilization, advertising that developers may optimize layouts, improve operational efficiency, and better meet the evolving needs of student residents, although such tools pose privacy risks \cite{fyma2025studentaccommodation}. Similarly, housing directors increasingly turn to platforms like Campus Labs Engage and Eventbrite for tracking attendance via geolocation, QR codes, and check-in apps \cite{campuslabs2025engage,scandit2025highereducation,eventbrite2025checkin}. However, tools like facial recognition, such as those offered by Trueface, may be seen as intrusive and risk compromising student privacy \cite{trueface-a}.

Career services departments are also leveraging AI to enhance offerings. Platforms like Handshake and Symplicity use machine learning to personalize job recommendations, improve search results, and connect students with employers. Handshake’s AI career copilot, Coco, is supposed to assist with interview preparation, while Symplicity’s Career Services Manager advertises a tailored user experience based on behavioral data \cite{handshake2025coco,symplicity2025careercentres}. While these tools may provide significant benefits, such as 24/7 access to career coaching and improved efficiency, they also carry risks.  To mitigate these risks of algorithms trained on WEIRD data and containing diverse and contradictory values, universities must carefully evaluate and monitor AI tools, prioritizing their vision of fairness, transparency, and accountability.

With numerous tools emerging—many of which will quickly become obsolete—universities must adopt ethical practices supported by effective oversight to ensure these technologies enhance accessibility, operational efficiency, and equitable outcomes \cite{smith2024a}. For example, university counseling and retention services increasingly use AI to enhance efficiency and student support while addressing ethical considerations and privacy concerns. Institutions like Ivy Tech Community College and Furman University claim this potential; Ivy Tech analyzes performance data to identify at-risk students for timely intervention, while Furman University enhances student well-being through an AI-powered personalized support app \cite{brady2024a}. Tools like AI scheduling assistants, such as those offered by Spring Health and TheraNest, are supposed to streamline appointment booking, improve accessibility, and reduce staff workload \cite{health-a,educause2025retention}.

Furthermore, retention coordinators employ AI-powered early-warning systems, like EAB Navigate and Starfish, to identify at-risk students through predictive analytics, enabling timely interventions. However, these systems carry risks of false positives or negatives \cite{eab2025navigate360,bauman2024textingrisk,eab2025aiandrecruitment,eab2025starfish,uoc2023AIrisk}. Similarly, AI chatbots like Woebot and Ivy.ai handle non-urgent mental health FAQs and provide 24/7 support, but their inability to detect complex emotional nuances underscores the need for human oversight \cite{woebothealth2025mentalhealth,ivyai2025studentcommunication,ellucian2025ivyai}. While organizations such as the Association for University and College Counseling Center Directors (AUCCCD) have said little about AI so far, the American Counseling Association (ACA) stresses the importance of ethical guidelines, including informed consent, data privacy, and ensuring that AI tools supplement rather than replace human interaction \cite{american2025technologycounseling,aucccd2025bestpractices}. 

Using a fragmented set of tools, many incorporating AI, presents significant challenges for student affairs departments similar to those in academic affairs. A lack of integration can result in inefficiencies such as duplicate data entry, inconsistent record-keeping, and difficulties tracking students across multiple systems \cite{e2023a}. This fragmentation often leads to siloed information, where critical data needed to support students holistically is scattered across unconnected programs \cite{brown2000a}. For instance, a retention tool might flag students as at-risk based on attendance patterns without considering career services data showing strong internship engagement. Fragmented systems also heighten privacy risks by storing sensitive student data across multiple platforms, increasing the chances of breaches or mismanagement \cite{educause2025dataprivacy,ecampusnews2025datauseai}. Additionally, students may feel frustrated by disjointed services, while staff struggle with disconnected systems, ultimately hindering personalized support.

To address these issues, student affairs departments should prioritize adopting integrated platforms or invest in middleware software to connect existing tools. Unified platforms that consolidate functions like retention tracking, housing, career services, and student engagement can streamline workflows and centralize critical data. Middleware solutions such as APIs, iPaaS, and ESBs facilitate seamless data sharing and system integration: APIs enable direct communication between systems, iPaaS simplifies workflows, and ESBs manage complex, enterprise-level interactions \cite{laserfiche2025summarization}. For example, middleware using APIs can aggregate data from various sources into a centralized dashboard, allowing staff to address student needs proactively \cite{ibm2025middleware}. However, relying on sensitive indicators beyond faculty-reported grades or attendance—such as data from campus jobs or security—raises significant privacy concerns in "at-risk" flagging systems. While middleware connects and standardizes data across systems, AI agents autonomously perform tasks, make decisions, and generate insights, often relying on middleware for the data necessary to power advanced operations like machine learning \cite{guran2024a}. Institutions must demand transparency from middleware and AI agent providers, establish robust data governance policies, audit system performance, and train staff on ethical AI use to ensure these tools enhance student outcomes while avoiding inefficiencies, inequities, or privacy violations.

In navigating AI's transformative potential, student affairs departments must adopt a balanced approach that embraces innovation while safeguarding ethical standards and human connections. NASPA, a leading authority for student affairs professionals, underscores the importance of integrating AI thoughtfully to uphold institutional values and support student success. As their recent report highlights, “AI should be viewed not as a replacement for student affairs professionals but as a powerful tool that enhances their capabilities.”\cite{brady2024a}  Institutions can build a more equitable and effective support system by strategically leveraging AI to streamline processes, enhance data-driven decision-making, and proactively address student needs. However, achieving this requires transparent governance, ongoing training, and a steadfast commitment to centering human interactions within AI-driven systems. As NASPA emphasizes, the future of student affairs lies in fostering “a powerful synergy” between technology and human expertise, ensuring that AI amplifies the mission of holistic student development, a liberal arts college trademark \cite{brady2024a}. 

Student affairs departments must act as stewards of both innovation and caution. They must adopt AI tools transparently, assemble interdisciplinary task forces to assess their impacts, and regularly review policies to protect student privacy and equity. Just as vital, they need to fix fragmented, disconnected systems that can undermine the very support they aim to provide. By treating students as partners in these efforts and anchoring decisions in their values, colleges can integrate AI in ways that enhance outcomes without losing sight of their higher mission. This is not just about implementing technology; it’s about shaping a future that balances innovation with the enduring need for human connection.

\subsection{AI, Legal Risks, and the Liberal Arts College Advantage}

The growing integration of AI tools into academic and student affairs offers opportunities for liberal arts colleges to advance their human-centered missions, but it also introduces significant legal risks, particularly in the areas of privacy compliance and protection against discrimination. The Family Educational Rights and Privacy Act (FERPA) grants students the right to access, amend, and interpret their education records, requiring that grades and evaluations be transparent and secure. However, the opaque “black box” nature of many AI systems complicates compliance. For instance, faculty using generative AI (GenAI) tools like ChatGPT to evaluate student work—such as essays or qualitative assignments—may save time and provide detailed feedback but risk violating FERPA if they cannot explain how grades or feedback were determined.\footnote{Some leading companies are beginning to address this challenge by introducing features that reveal the reasoning behind generated content or predictions.}\cite{education2020a,usdepteducation2021,mckinsey2025,openai2025}  This challenge is particularly acute in subjective assessments, such as evaluating poetry, where a GenAI tool might penalize creative choices—like using the color "blue" to evoke melancholy or employing a traditional sonnet form—by labeling them as "conventional" or "unoriginal." Similarly, culturally rich or vernacular elements from Haitian Creole or Vietnamese American students might be misinterpreted as "errors" due to the tool’s training data, which often reflect WEIRD norms. Without faculty oversight to address these limitations, AI-generated evaluations risk unfairly disadvantaging certain groups, raising concerns about inclusivity, and potentially leading to legal challenges under FERPA or Title IX \cite{jacques2024a}.

FERPA’s requirement that grades and evaluations be interpretable, accessible, amendable, and secure while protecting personally identifiable information (PII) creates significant challenges in the fragmented landscape of AI tools with its inconsistent privacy and security practices.\cite{education2020a} Although leading GenAI companies like OpenAI and Google have adopted measures such as encryption, enterprise-grade security, and techniques like differential privacy and data anonymization, their implementation remains inconsistent across the industry \cite{openai-b,openai-a,google2025,google-b,golda2024a,yao2024a}.\footnote{Differential privacy, for instance, attempts to prevent reverse-engineering individual data points from model outputs, protecting user privacy during training \cite{yan-a}.}  Anthropic’s Claude is marketed as a more secure large language model (LLM) due to its “Constitutional AI” approach, but its claims require further evaluation \cite{anthropic-a}. Open-source models like Meta’s LLaMA and China’s DeepSeek present additional challenges because their decentralized nature places responsibility for privacy on individual developers, increasing the risk of misuse or inadequate safeguards \cite{meta-a}. A shared concern among security experts is the inadvertent exposure of sensitive data, particularly when student PII is logged for model improvement \cite{unknown-a}.\footnote{According to the company, “Google collects your Gemini Apps conversations, related product usage information, info about your location, and your feedback. […] Please don’t enter confidential information in your conversations or any data you wouldn’t want a reviewer to see or Google to use to improve our products, services, and machine-learning technologies.”\cite{google-a}} While no significant breaches involving leading GenAI companies have been reported, the prevalence of corporate data breaches suggests such an event is likely \cite{security-a}. These discrepancies underscore the urgent need for greater transparency, standardization, and government regulation to ensure robust and equitable privacy protections across AI platforms \cite{michael2025ai}. In the AI arms race with billions of dollars at stake, these companies are unlikely to be more forthcoming or advance industry cooperation without additional government oversight.

Some states have enacted their own legislation that affects the legality of the use of GenAI in university settings.  California legislation like AB 1584 (enacted in 2014) directly addresses many concerns about data privacy in educational settings, requiring that data shared with third parties remains the property of local educational agencies.  Student data "shall not be used by the third party for any purpose other than those required or specifically permitted by the contract." It also mandates strict security and confidentiality measures and requires vendors to notify educational agencies of unauthorized disclosures. Theoretically, this framework should regulate AI tools, ensuring that data shared for personalizing learning or assessments is strictly controlled, but the onus seems to be put on users currently. Coupled with broader laws like the California Consumer Privacy Act (CCPA), enacted in 2018 to enhance individual control over personal data, and the Children's Data Privacy Act, introduced in January 2024 to strengthen protections for minors, AB 1584 provides a robust legal foundation to mitigate data misuse and breaches. Despite these regulations, many California universities may inadvertently violate them through the unregulated use of GenAI and PAI applications. 

Considering universities' risk of exposing themselves to legal liabilities while undermining institutional accountability and student trust, it is no wonder that some of the bigger and better-financed universities are turning toward their own proprietary GenAI systems \cite{unknown2024a}. For instance, the University of Michigan has developed U-M GPT to address security and ethical concerns, while the University of California, Irvine has introduced ZotGPT, a “secure” AI platform with UCI-specific data search capabilities \cite{university2024generative,university2024a}. While such tools offer enhanced privacy and customization, they often use the same WEIRD training data and may be as, or even more, vulnerable to breaches. Additionally, many proprietary GPTs have limited or no internet access, leading staff and students to turn to unregulated or unpermitted tools that access up-to-date data or information.

Beyond privacy and security, the GenAI’s WEIRD training data may pose a unique risk in higher education. Student affairs staff using red-flag PAI systems or GenAI to help with selection processes (e.g., housing, clubs, jobs) might unfairly penalize certain students, particularly those from underrepresented or international groups. This concern is evident in the COMPAS legal case, where an AI risk assessment tool used in the criminal justice system disproportionately flagged Black defendants as having a higher risk for recidivism compared to White defendants. Developed by Northpointe (now Equivant), COMPAS relied on historical data and opaque algorithms, resulting in public criticism and calls for transparency after \textit{ProPublica} revealed its flaws \cite{angwin2016machine,suresh2021framework}. Some argue, however, that \textit{ProPublica} misinterpreted key statistical principles and ignored the broader context of risk assessment in criminal justice \cite{flores2016a}. While COMPAS used predictive AI, the underlying issue of pluralism and value-restrictive training data is equally relevant to generative AI systems, which may inadvertently perpetuate representational harms through stereotypes or cultural blind spots in student-related decision-making processes.

A similar pattern of controversy emerged with Proctorio, an AI-powered exam monitoring tool adopted widely during the COVID-19 pandemic. Proctorio has been accused of using invasive and algorithmic tracking technologies, such as webcams and keystroke monitoring, which may disproportionately affect students with disabilities, lower-income students, and those with darker skin tones due to facial recognition biases \cite{oliver2021reckoning,cdt2025proctoring,cox2021proctorio}. While Proctorio relied on predictive AI to flag "suspicious" behaviors, generative AI poses related risks in higher education, such as generating misleading content or responses influenced by its biased training data. Proctorio has acknowledged these concerns, citing third-party audits that found no significant bias, and is working to improve its software for fairness and inclusivity \cite{proctorio-a}. Similarly, Wells Fargo’s AI-driven mortgage lending system, which disproportionately denied loans to minority applicants, highlights how systemic inequities embedded in training data can lead to discriminatory outcomes \cite{donnan2022wells}. these examples underscore the necessity for rigorous oversight and fairness auditing to prevent AI-related harms in education. When carefully designed and monitored, AI—whether predictive or generative—might reduce human errors and biases, enhancing equity and consistency in decision-making processes.

Open-source AI models like Meta’s LLaMA and High-Flyer’s DeepSeek-V3, along with proprietary systems like those developed by the University of Michigan and UC Irvine, illustrate both the potential for democratizing AI adoption and the pressing need for comprehensive standards and oversight to address the legal, ethical, and privacy challenges universities face. Open-source models reduce dependency on proprietary systems and may offer institutions greater control over data privacy, provided they can effectively mitigate breaches or attacks. The potential for enhanced customization and independence may justify the additional effort required to manage these systems. However, the coexistence of open-source and proprietary models underscores a fragmented and inconsistent approach to AI governance in higher education. This inconsistency highlights the critical need for comprehensive oversight to ensure transparency, accountability, and equity in AI implementation. University accreditation agencies, as arbiters of institutional accountability and quality, are uniquely positioned to guide the ethical and responsible integration of AI into academic and student affairs, helping institutions navigate these opportunities and risks effectively.

\section{AI and University Accreditation}

Considering the rapid adoption of AI and its associated ethical and legal challenges, it is no surprise that university accrediting agencies have begun to address the issue. However, their approaches remain inconsistent and largely preliminary. Some, like the Southern Association of Colleges and Schools Commission on Colleges (SACSCOC), have issued concrete guidelines emphasizing confidentiality, data security, and the risks of over-relying on AI for accreditation materials \cite{sacscoc2024ai}. Similarly, the Higher Learning Commission (HLC) has acknowledged AI’s potential for efficiency while cautioning against risks like bias and academic integrity violations. By contrast, agencies like the Middle States Commission on Higher Education (MSCHE) and the New England Commission of Higher Education (NECHE) have limited their engagement to webinars and discussions without issuing formal policies. Meanwhile, the Western Association of Schools and Colleges Senior College and University Commission (WSCUC) and the Northwest Commission on Colleges and Universities (NWCCU) have focused on ethical principles, such as transparency and substantive instructor-student interaction, but have not provided specific directives for generative AI. This variation leaves institutions, particularly LACs, with uneven guidance on how to integrate AI responsibly.

The few initial responses from accrediting agencies reveal significant differences in approach and depth. HLC, while acknowledging AI's potential and risks, has focused more on raising awareness and seeking information through surveys and reports, offering fewer actionable strategies \cite{unknown-b}. Agencies like MSCHE and NECHE have not moved beyond exploratory discussions, leaving their member institutions without specific guidelines for addressing generative AI’s challenges. In contrast, NWCCU and WSCUC emphasize ethical considerations, such as transparency and interaction standards, which align with the values of mission-driven institutions like LACs but fall short of addressing the technical and operational complexities of AI integration \cite{nwccu2025ai,wscuc2024policy}. Overall, the responses vary widely, with some agencies offering pragmatic advice and others limiting their engagement to general principles or exploratory events, creating a patchwork of guidance that complicates AI adoption for smaller, resource-constrained institutions.

The SACSCOC "Artificial Intelligence in Accreditation" document and WSCUC’s draft "Artificial Intelligence Limits and Peer Review of Institutional Reports" policy highlight accrediting bodies' cautious approaches toward AI integration. SACSCOC emphasizes security and confidentiality while warning against overreliance on generative AI, though its broad risk generalizations and limited actionable strategies could hinder innovation \cite{sacscoc2024ai}. In contrast, WSCUC prohibits external AI tools in peer review but allows for Commission-approved AI, perhaps trying to balance security concerns with modernization \cite{wscuc2024ai}. Both policies reflect a commitment to integrity and ethical use. Still, they would benefit from clearer differentiation between AI types, actionable guidelines beyond peer review reports, and support for low-risk, beneficial applications to help institutions navigate AI integration responsibly. Additionally, the WSCUC’s suggestion for the Commission-approved AI again points to the fragmentation of AI tools, each tailored for institutional needs and to guard against security risks.  

The hesitation and limitations shown by regional accreditation agencies in addressing AI applications reflect broader challenges in adapting to emerging technologies. Historically, these agencies have emphasized integrating technology to enhance educational quality, accessibility, and institutional effectiveness \cite{unknown2021a,s2024b,hlc_criteria_2025}. Building on this foundation, future accrediting guidelines will likely require institutions to adopt policies that prevent academic dishonesty, ensure data privacy and security, and promote professional development to help faculty and staff ethically integrate AI. Equity and accessibility will remain central, encouraging AI implementations that benefit all students. Institutions will also need to assess AI’s impact on learning outcomes and operations continuously, fostering adaptability to technological change. These efforts would align with accrediting agencies’ missions to uphold educational quality and institutional integrity in an increasingly AI-driven world.

We should expect controversies and lawsuits, such as data breaches exposing student information through AI vendors or widespread cheating facilitated by AI tools, to test accrediting agencies’ roles in enforcing federally mandated expectations. While not directly liable, accrediting bodies act as intermediaries between institutions and federal regulators, and systemic failures in areas like data protection or academic honesty could jeopardize their recognition by the U.S. Department of Education. For example, a data breach could reveal gaps in vendor management under laws like FERPA, while cheating scandals might highlight inadequate safeguards against academic dishonesty. To address these issues, agencies must refine their guidelines, emphasizing robust risk management protocols and updated standards for AI misuse. Beyond hosting webinars, agencies should engage more directly and publicly with institutions, as HLC did through its published needs survey, to strengthen oversight and adapt to the challenges of an increasingly AI-driven educational landscape \cite{commission-b}.

In addition to domestic concerns, the global nature of higher education introduces challenges in navigating differing AI regulatory frameworks. Institutions collaborating internationally may face conflicting standards, such as Europe’s stringent AI Act, versus regions with more lenient or undefined policies \cite{commission-a,nodes-a}. This misalignment complicates international partnerships and may impose burdens on institutions trying to comply with multiple regulations. Accrediting agencies have yet to provide substantial guidance on reconciling these disparities, leaving universities vulnerable to inefficiencies and missed opportunities. Agencies should prioritize regular policy updates and foster collaboration with global partners to ensure AI integration aligns with evolving technologies and diverse regulatory requirements.

\section{Conclusion}

In the coming years, liberal arts colleges will face a defining moment in determining how artificial intelligence reshapes their institutional identity, pedagogy, and operations. AI offers opportunities to enhance efficiency, streamline administrative burdens, and expand student support services, yet it also presents profound ethical, legal, and philosophical challenges that demand deliberate governance. The fragmented and often contradictory approaches to AI regulation, value alignment, and fairness demonstrate that no universal solution exists—only mission-driven, institutionally grounded strategies. LACs can harness their close-knit, mission-driven communities to integrate AI thoughtfully, ensuring it aligns with their values of interdisciplinary learning, inquiry, and student-centered education. However, financial and technological constraints pose hurdles, particularly as the rapid proliferation of AI tools risks exacerbating institutional disparities. The growing competitiveness of open-source models marks a pivotal moment for LACs, potentially offering a way to overcome steep financial barriers, adopt AI in ways that reflect their educational missions, and uphold equity and inclusivity—provided these tools remain safe and secure.

The path forward is not simply about whether to adopt AI but how to do so in a way that reinforces LACs’ core commitments to holistic education and student-centered learning. To navigate this complexity, LACs must proactively integrate AI into their institutional frameworks while maintaining faculty oversight, ethical safeguards, and a commitment to human judgment in decision-making. By recognizing that values cannot be “aligned” except within united communities, they can approach this technology more intentionally and safely. Effective AI adoption should support, rather than replace, faculty expertise and human relationships that define a liberal arts education. Additionally, strategic investments in technology, partnerships with research universities, and robust governance structures will be critical to ensuring that AI systems reflect the diverse values of their communities while advancing transparency and accountability. By treating AI not as a force to be passively managed but as a tool to be actively shaped in service of their mission, LACs can lead higher education in developing a model of AI adoption that is both innovative and ethically responsible—one that upholds the transformative power of human learning even in an era of machine intelligence.

Placing mission and values first puts AI tools to work for the university, not the other way around. For example, universities are formulating their position on AI use.  We might again prompt a popular GenAI tool to create these positions for the same fictitious universities used as examples in the introduction, highlighting value divergence in AI output. When the tool was asked to craft a guiding principle and action statements regarding AI use for a progressive, globally oriented West Coast college and a conservative, Christian-driven institution in the South, a remarkable contrast emerged, as revealed in the following table.

\begin{table}[H]
\centering
\caption{AI-Generated Guiding Principles and Action Statements for Two Hypothetical LACs}
\label{tab:3}
\begin{adjustbox}{max width=\linewidth}
\begin{tabular}{L{3cm}L{5cm}L{7.5cm}}
\toprule
\textbf{Institution}                                               & \textbf{Guiding   Principle}                                                                                                                                                                                                                                                                                                     & \textbf{Action   Statements}                                                                                                                                         \\
\midrule
\multirow{3}{3cm}{Conservative, Christian-driven LAC in   the South} & \multirow{3}{5cm}{“We are dedicated to the   ethical development and application of artificial intelligence, ensuring its   use upholds biblical principles, respects human dignity, and serves to   advance Christlike character and stewardship in all endeavors.”}                                                              & “To apply AI in education   and administration in ways that reflect and reinforce biblical principles and   moral integrity.”                                        \\
                                                                   &                                                                                                                                                                                                                                                                                                                                  & “To ensure that AI systems   respect human dignity, safeguard privacy, and promote stewardship as part of   God’s creation.”                                         \\
                                                                   &                                                                                                                                                                                                                                                                                                                                  & “To equip students with   the discernment to ethically engage with AI technologies, using them to   advance Christlike service and gospel proclamation.”             \\
\multirow{3}{3cm}{Progressive,   globally oriented West Coast LAC}   & \multirow{3}{5cm}{“We are committed to   advancing AI’s ethical and human-centered use in education, ensuring it   fosters creativity, critical thinking, and inclusivity while safeguarding   human dignity, equity, and trust in alignment with our mission to create   global citizens dedicated to peace and sustainability.”} & “To leverage AI to   enhance creativity, critical thinking, and inclusivity, while aligning with   principles of human rights, dignity and sustainability.”          \\
                                                                   &                                                                                                                                                                                                                                                                                                                                  & “To ensure transparency,   ethical governance, and accountability in AI applications within educational   and administrative contexts.”                              \\
                                                                   &                                                                                                                                                                                                                                                                                                                                  & “To educate students on   the responsible use of AI as a tool for global citizenship and collaborative   problem-solving for societal and environmental challenges.” \\
                                                                   \bottomrule
\multicolumn{3}{p{\linewidth}}{ChatGPT, response to “Create a single guiding principle and three action statements for ethical AI use for a conservative, Christian-driven LAC in the South,” OpenAI, January 3, 2025; ChatGPT, response to “Create a single guiding principle and three action statements for ethical AI use for a progressive, globally oriented LAC on the West Coast,” OpenAI, January 3, 2025.}
\end{tabular}
\end{adjustbox}
\end{table}

These principles on AI usage are more than lofty statements or removed from pedagogy and student life within the university. For example, the West Coast college emphasized inclusivity, sustainability, and critical thinking, positioning AI as a tool for advancing global citizenship and human rights and solving environmental challenges. Its principles reflected a commitment to fostering creativity and equity while ensuring transparency and ethical governance in AI systems. In contrast, the Southern Christian institution’s statements rooted ethical AI use in biblical values, prioritizing moral integrity, human dignity, and stewardship under a Christian worldview. Action items included using AI to enhance spiritual growth, safeguarding data privacy while respecting God’s creation, and avoiding technologies conflicting with scriptural teachings. These examples reveal how institutional missions and their foundational beliefs can lead to starkly different approaches to “ethical” AI, even as both seek to use technology responsibly and purposefully in their educational contexts. This distinction illustrates that AI principles are not merely aspirational but deeply connected to a university's pedagogy, resource allocation, and the framing of contentious topics within the curriculum.

As Ethan Mollick aptly notes, “Today’s AI is the worst AI you will ever use.”\cite{mollick2024a} The next decade will likely bring innovations that reshape nearly every university function, requiring intentional oversight and iterative adaptation. Liberal arts colleges, with their interdisciplinary ethos and human-centered missions, are well-positioned to lead in this area. However, success will require strategic investments in technology, partnerships with research universities, careful training of faculty and staff, and robust governance to ensure that AI systems reflect the diverse values of their communities while advancing equity and accountability \footnote{A good reference for this process is the EDUCAUSE Action Plan: AI Policies and Guidelines, which provides strategic recommendations and actionable steps for higher education institutions to develop effective AI policies, particularly in addressing the ethical and legal implications of AI technologies. While it emphasizes collaboration, transparency, and continuous adaptation to navigate AI’s evolving challenges and opportunities in academia, the report does not sufficiently account for how GenAI is both value pluralistic and contradictory, reflecting the biases and constraints of its WEIRD training data. Therefore, universities should begin by designing a mission-aligned AI strategy and use that alignment as the benchmark for measuring progress \cite{e2024a}.}. In doing so, these institutions can demonstrate how technology can complement rather than compromise the transformative power of education when guided by human judgment.

\bibliographystyle{unsrt} 
\bibliography{main}

\end{document}